\documentclass[aps,prb,twocolumn,floatfix]{revtex4-1}

\usepackage[dvips]{graphicx}
\usepackage{epsfig}
\usepackage{float,epsfig}
\usepackage{dcolumn}
\usepackage{bm}
\usepackage{amsmath}
\usepackage{ulem}

\begin{document}
\title {Overlapping fragments method for electronic structure calculation of large systems}

\author{Nenad~Vukmirovi\'c$^{1,2,a)}$\footnote[0]{$^{a)}$ Electronic mail:  nenad.vukmirovic@ipb.ac.rs} 
 and Lin-Wang Wang$^1$}
\affiliation{$^1$Materials Sciences Division, Lawrence Berkeley National
    Laboratory, Berkeley, CA 94720, USA.\\
$^2$Scientific Computing Laboratory, Institute of Physics Belgrade, University of Belgrade, Pregrevica 118, 11080 Belgrade, Serbia}

\begin{abstract}
We present a method for the calculation of electronic structure of systems that contain tens of thousands of atoms. The method is based on the division of the system into mutually overlapping fragments and the representation of the single-particle Hamiltonian in the basis of eigenstates of these fragments. In practice, for the range of system size that we studied (up to tens of thousands of atoms), {the dominant part of the calculation scales} linearly with the size of the system when all the states within a fixed energy interval are required. The method is highly suitable for making good use of parallel computing architectures. We illustrate the method by applying it to diagonalize the single-particle Hamiltonian obtained using the density functional theory based charge patching method in the case of amorphous alkane and polythiophene polymers. 
\end{abstract}


\maketitle

\section{Introduction}
In the last few decades, density functional theory (DFT)~\cite{PhysRev.136.B864} became a method of choice for the calculation of the electronic structure of physical systems with a relatively large  number (hundreds to about a thousand) of atoms. Within DFT, one has to self-consistently solve the Kohn-Sham equations~\cite{PhysRev.140.A1133} for the wave functions $\psi_i$ and energies $\varepsilon_i$
\begin{equation}
\left(-\frac{\hbar^2}{2m_0}\nabla^2+V_\mathrm{ion}+V_\mathrm{H}+V_\mathrm{xc}\right)\psi_i=
\varepsilon_i\psi_i, \label{eq:eqA1}
\end{equation}
where
$V_\mathrm{ion}$ is the potential of the core ions,
$V_\mathrm{H}$ is the electrostatic (Hartree) potential of the electronic charge density distribution $\rho({\bm r})$
and 
$V_\mathrm{xc}$ is the exchange correlation potential which, under the local density approximation (LDA), depends only on charge density at a given point in space. 

There is a strong interest to develop methods where the cost of solving the system of equations (\ref{eq:eqA1}) would depend linearly on the number of atoms in the system. Such methods are based either on the representation of DFT equations in localized orbital basis sets \cite{prb64-235111,jcp122-084119} or on the division of the system into small fragments.\cite{cpl313-701,prb77-165113} These methods are still computationally demanding due to necessity of evaluating all the wave functions of occupied states in each iteration until the self-consistency is reached.

A different class of (empirical) methods has been developed in the semiconductor physics community, where the main philosophy is to directly construct the Kohn-Sham Hamiltonian [the left hand side of Eq.~(\ref{eq:eqA1})]. In the empirical and semiempirical pseudopotential method (EPM and SEPM), the total potential is considered as a sum of pseudopotentials of individual atoms, that are obtained either by fitting to the bandstructure of a bulk semiconductor \cite{PhysRev.141.789,PhysRevB.14.556} or extracted from ab-initio calculations of the bulk.~\cite{prb51-17398} Such pseudopotentials are then used to construct the Hamiltonian of the nanostructure of interest.~\cite{prb51-17398,prb55-1642} A more recent approach is the charge patching method (CPM),\cite{prl88-256402,jcp128-121102} where the electronic charge density is constructed from charge density contributions of individual atoms -- so called motifs. The motifs are extracted from calculations on small prototype systems, where the atoms have a similar bonding environment as in the system of interest. For a range of inorganic and organic semiconducting systems,\cite{prl88-256402,prb67-033102,prb67-205319,prb72-125325,jcp128-121102,nl-un,jpcb113-409} the charge density and the potential obtained from the CPM closely match the ones that would be obtained from a full self-consistent DFT calculation. The construction of the Hamiltonian in the methods mentioned above (EPM, SEPM and CPM) is quick and its cost scales linearly with the size of the system.

Once the Kohn-Sham Hamiltonian is constructed, one has to solve its eigenvalue problem. For semiconducting systems, the spectral region of interest is the one in the vicinity of the band gap. Therefore, one needs to solve for these electronic states only. This can be achieved using the folded spectrum method.\cite{jcp100-2394} The folded spectrum method (implemented in plane wave representation of the wave functions) scales linearly with the size of the system when a fixed number of states is required. However, in many calculations, one is interested in a fixed energy window of the order of several $k_\mathrm{B}T$ below or above the band gap, since this is the spectral region that determines the electronic transport properties of the system. The number of states in such an energy window also increases linearly with the size of the system and consequently the overall computational cost within the folded spectrum method increases quadratically with the system size. 

In this paper, we present a different strategy for the diagonalization of the Hamiltonian. It is based on the idea of representing the Hamiltonian in a localized and physically well motivated basis. The whole system is divided into many small fragments, that are not necessarily disjoint, and the eigenstates of the fragments are chosen as the basis for the representation of the Hamiltonian. In some sense, this approach combines the ideas from the literature on using the localized basis sets and the division of the system into fragments. We will refer to this method as overlapping fragments method (OFM).

We have developed this methodology with a particular focus towards its application to understanding the electronic states in semiconducting polymer materials. These materials are to a large extent disordered and there is a strong need for large supercell calculations that would provide reliable information about the degree of localization of electronic states, the density of states and eventually the electronic transport properties.\cite{nl-un,prb81-035210,apl97-043305} For such systems, the atomic structure can be reliably generated from classical molecular dynamics (MD),\cite{chemphyschem5-373,acsn2-1381,prb76-241201,cpl480-210,jcp132-134103,jacs131-11179,jpcb113-409,jpcb108-6988,jpcb111-11670} 
but the challenge remains to calculate the electronic structure.
The current method excellently complements our recently developed CPM for the construction of the Hamiltonian of organic semiconducting materials.

We present the details of the implementation of the methodology in Sec.~\ref{sec:sec2}. In Sec.~\ref{sec:sec3}, we illustrate the method to diagonalize the Hamiltonian obtained from CPM in the case of alkanes and describe the main points that one should address when performing such a calculation. Finally, in Sec.~\ref{sec:sec4}, the method is illustrated by an application to one of the most widely studied organic polymers -- poly(3-hexylthiophene) (P3HT).

\section{Method and Implementation}\label{sec:sec2}

\begin{figure*}[htbp]
 \centering
  \epsfig{file=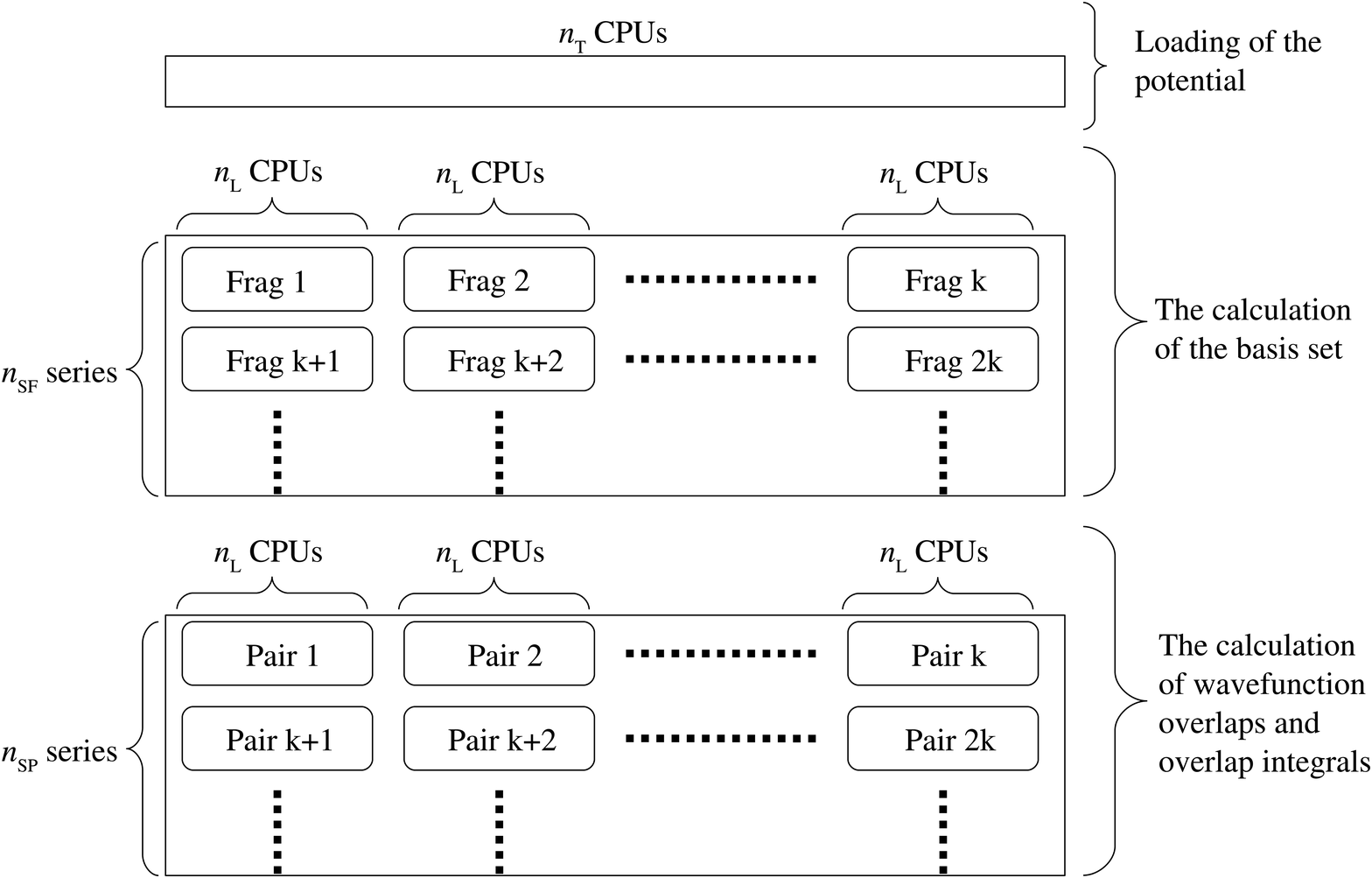,width=15cm,angle=0}
  \caption{The scheme describing the implementation of the OFM on parallel computers.}
 \label{fig:figA1}
 \end{figure*}

In this section we describe the details of the OFM and its implementation on parallel computers. The input to our computation is the atomic structure of the system and its potential obtained from CPM, while the output gives the transfer integrals 
$H_{ij,mn}=\langle \phi_i^{(j)}|H|\phi_m^{(n)}\rangle$ 
and the wave function overlaps 
$S_{ij,mn}=\langle \phi_i^{(j)}|\phi_m^{(n)}\rangle$
between the pairs of states $\phi_i^{(j)}$ and $\phi_m^{(n)}$, which are the $i$-th wave function of the fragment $j$ and the $m$-th wave function of the fragment $n$. Each fragment consists of a molecule (its choice will be discussed later) embedded in a cuboid box. The potential is stored on all the $n_\mathrm{T}$ CPUs available for computation, as schematically illustrated in Fig.~\ref{fig:figA1}. 

The computation consists of two main parts (Fig.~\ref{fig:figA1}): the calculation of basis wave functions and the calculation of $H_{ij,mn}$ and $S_{ij,mn}$.
We allocate $n_\mathrm{L}$ CPUs to each of the fragments, where $n_\mathrm{L}$ is typically some small number (for example $n_\mathrm{L}=8$ or $16$). 
The calculation of the basis wave functions stemming from a given fragment is performed as follows. The charge density of the fragment is obtained using the charge patching method by adding the charge density motifs of each of the atoms in the fragment. The Hartree potential of the fragment is then evaluated from the solution of the Poisson equation with periodic boundary conditions, and the exchange correlation potential is obtained from the LDA formula. In such a way, one obtains the Hamiltonian of the fragment, which is then diagonalized using the ESCAN code,\cite{jcp160-29} which implements the preconditioned conjugated gradient minimization algorithm with the plane wave basis set. The basis wave functions $\phi_i^{(j)}$ stemming from each fragment $j$ are therefore obtained. At this stage, we also calculate $H|\phi_i^{(j)}\rangle$ (where $H$ is the Kohn-Sham Hamiltonian of the whole system, not just the fragment), which will be later required for the evaluation of $H_{ij,mn}$. To achieve this, 
one first has to send the required real space grid values of the potential to the $n_L$ CPUs allocated for fragment $j$. The $H|\phi_i^{(j)}\rangle$ operation is then performed using one of the main subroutines from the ESCAN code. One should note that the speed of the calculation of basis wavefunctions can be further improved by using some localized basis set instead of plane waves.

Let $n_\mathrm{F}$ be the number of fragments in the system and 
$k=\lceil n_\mathrm{T}/ n_\mathrm{L} \rceil$. We divide the fragments into $n_\mathrm{SF}=\lceil n_\mathrm{F}/k \rceil$ series, as illustrated in Fig.~\ref{fig:figA1}. 
The calculations on the fragments from the same series are performed in parallel, where each fragment uses its $n_\mathrm{L}$ allocated CPUs (see Fig.~\ref{fig:figA1}). Such calculations are repeated for all $n_\mathrm{SF}$ series of fragments. 

The code for performing the above tasks was written by making good use of the existing codes for performing the charge patching calculations, solving the Poisson equation and the ESCAN code. These were integrated into a single code to avoid reading and writing to disk of the input and output files, such as charge densities and potentials, which can be quite large. For the storage of the calculated basis wave functions $\phi_i^{(j)}$, their reciprocal space representation is used. Each fragment will have only a few basis functions, as discussed below, and therefore the required memory for their storage is not very big. Each of the existing codes has been already parallelized using Message Passing Interface (MPI). Further parallelization with respect to fragments, described above, was achieved by using the \verb|mpi_split| command and changing the \verb|mpi_comm_world| communicator in the existing codes to a local communicator (among the $n_\mathrm{L}$ CPUs) defined by the \verb|mpi_split| command.

The main part of the calculation consists of the calculation of the transfer integrals 
$H_{ij,mn}$ 
and the wave function overlaps 
$S_{ij,mn}$ 
between the pairs of states $\phi_i^{(j)}$ and $\phi_m^{(n)}$. Since the wave function $\phi_i^{(j)}$ is well localized to the fragment $j$, one naturally introduces an approximation to consider only the transfer integrals and wave function overlaps for the states $\phi_i^{(j)}$ and $\phi_m^{(n)}$, such that the fragments $j$ and $n$ are not too distant in space. The exact criterion for this will be formulated later in the paper. Let $n_\mathrm{P}$ be the number of pairs of fragments $\{j,n\}$ for which $H_{ij,mn}$ and $S_{ij,mn}$ need to be evaluated. For each pair, we allocate $n_\mathrm{L}$ CPUs where the calculation is performed. The pairs are divided into $n_\mathrm{SP}=\lceil n_\mathrm{P}/k \rceil$ series, in a similar manner as fragments (Fig.~\ref{fig:figA1}). To perform the calculation of $H_{ij,mn}$ and $S_{ij,mn}$ for pair $\{j,n\}$ on its allocated $n_\mathrm{L}$ CPUs, one needs to receive the wave functions $\phi_i^{(j)}$, $H\phi_i^{(j)}$, $\phi_m^{(n)}$ and $H\phi_m^{(n)}$, which are stored on different groups of $n_\mathrm{L}$ CPUs, the ones associated with fragments $j$ and $n$.
With the wave functions available on the allocated group of CPUs, the overlap element $S_{ij,mn}$ is straightforwardly calculated from the overlap of $\phi_i^{(j)}$ and $\phi_m^{(n)}$, while $H_{ij,mn}$ is calculated as the overlap of either $\phi_i^{(j)}$ and $H\phi_m^{(n)}$ or $H\phi_i^{(j)}$ and $\phi_m^{(n)}$.  In a similar manner as for fragments, the calculations for pairs from the same series are performed in parallel (Fig.~\ref{fig:figA1}), and then sequentially repeated for all $n_\mathrm{SP}$ series of pairs.

With $S_{ij,mn}$ and $H_{ij,mn}$ elements at hand, the final step is to find the electronic states by solving the generalized eigenvalue problem
\begin{equation}
\sum_{mn}\left(H_{ij,mn}-ES_{ij,mn}\right)C_{mn}=0.
\end{equation}
As will be shown, a limited number of basis wave functions is sufficient for rather accurate results. Therefore, the dimension of the matrices $S_{ij,mn}$ and $H_{ij,mn}$ is not very large. Consequently, in our current implementation of the methodology, this part is performed as a postprocessing step by using the standard LAPACK\cite{LAPACK} single processor routines. 
{For very large systems or basis sets, we use ScaLAPACK.\cite{ScaLAPACK} One can in principle also exploit the fact that the matrices $H_{ij,mn}$ and $S_{ij,mn}$ are sparse and use PARPACK\cite{PARPACK} which is well suited in that case.}

We note that a method exploiting to some extent similar ideas has been recently proposed by McMahon and Troisi,~\cite{cpl480-210} in the context of the calculation of electronic structure of semiconducting polymers. Their method is also based on the partitioning of the system into fragments and calculating the transfer integrals and basis wave function overlaps. In their method, the transfer integrals are however evaluated from the calculation of the system of two fragments in vacuum, which is inevitably an approximation. In our case, on the other hand, the transfer integrals are evaluated from the Kohn-Sham Hamiltonian of the whole system. Such transfer integrals therefore fully include all the other environmental factors surrounding the two fragments. In the case of disordered polymers we therefore do include the variations of both on-site energies and hopping integrals due to random electrostatic potential. 

\section{Example of the calculation: disordered alkanes}\label{sec:sec3}

In this section, we would like to illustrate the methodology by applying it to the alkane polymer system. As a test system, we choose 20 alkane chains, each one being 20 monomers long (1240 atoms altogether). We generate the atomic structure of the disordered chain from classical MD using a simulated annealing procedure. The CFF91 force field,\cite{jcc15-162,jacs116-2515} as implemented in the LAMMPS code\cite{lammps,jcp117-1} is used in the simulation.

\subsection{Choice of fragments}

\begin{figure}[htbp]
 \centering
  \epsfig{file=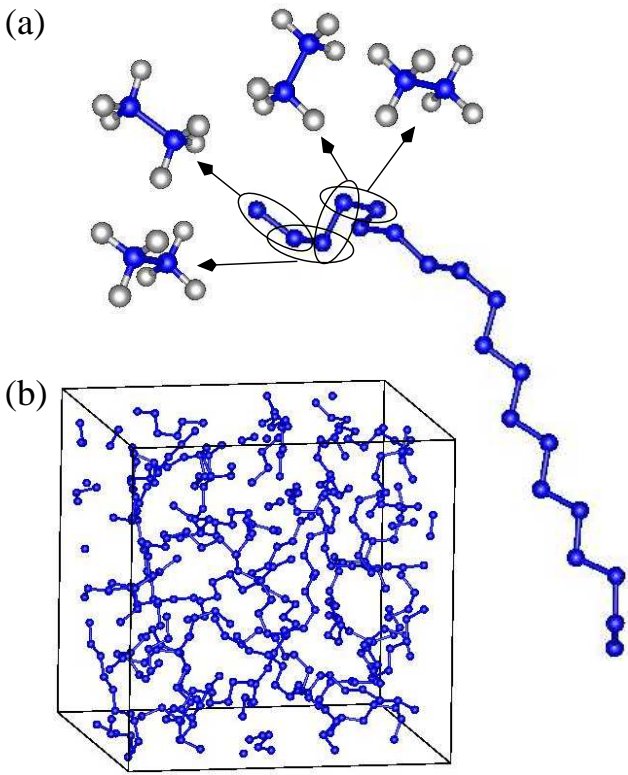,width=8.5cm,angle=0}
  \caption{(Color online) (a) A disordered chain of C$_{20}$H$_{42}$ and its division into overlapping dimer fragments. Several first fragments are shown only. (b) An amorphous system consisting of 20 C$_{20}$H$_{42}$ chains. Hydrogen atoms have been removed for clarity.}
 \label{fig:figA2}
 \end{figure}

The main task necessary to successfully apply the described methodology is to find the best way for the division of the system into fragments. A seemingly natural way is to cut the polymer into monomers, and to passivate the broken bond in each monomer by the hydrogen atom. In the case of alkanes, this would lead to the division of each $n$-units long alkane chain into $n$ CH$_4$ molecules. There is however a concern whether the basis set formed from the eigenstates of monomer fragments would be sufficient to reliably describe the wave function of the whole system, in particular in the region of the broken bond. A way around this problem is to choose a basis set formed from overlapping dimer fragments, illustrated in Fig.~\ref{fig:figA2}a. In such a way $n-$units long alkane chain is divided into $(n-1)$ C$_2$H$_6$ molecules. The main advantage of this way of the division of the system into fragments is that each bond in the system is fully encompassed by at least one fragment. We expect further improvement in the results when the system is divided into trimer fragments (C$_3$H$_8$ molecules). On the other hand, one should keep in mind that the choice of the fragments that are too large is not advantageous from the computational point of view, as the wave functions of all fragments need to be evaluated.

\begin{figure}[htbp]
 \centering
  \epsfig{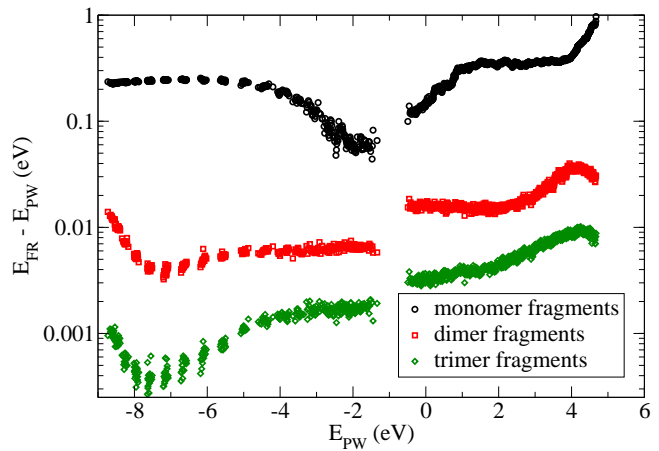}
  \caption{(Color online) The comparison of eigenenergies of amorphous alkane system calculated in the basis of fragment wave functions $E_\mathrm{FR}$ and plane waves $E_\mathrm{PW}$.  {The basis set consisting of all occupied states of the fragments was used in the calculation. All occupied states are shown in the figure. The Fermi level is at around 5~eV.}}
 \label{fig:comp_FR_PW}
 \end{figure}

To test the accuracy of the methodology, we have also solved the eigenvalue problem of the whole Hamiltonian in the plane wave basis set with kinetic energy cutoff of 60~Ry, using the ESCAN code. 
In Fig.~\ref{fig:comp_FR_PW}, we compare the eigenenergies {of all occupied states} obtained with the plane wave basis set $E_\mathrm{PW}$ and the eigenenergies obtained with the basis of fragment wave functions $E_\mathrm{FR}$. We include in the basis set all occupied states of the fragment, which constitutes four, seven and ten states for the cases of monomer, dimer and trimer fragment, respectively. As one might have expected, the results from using the basis of monomer fragments are not so accurate. In the lowest part of the spectrum, they are shifted by more than 200~meV from the results obtained in plane wave basis, while in the part of the spectrum near the top of the valence band the errors are of the order of 1~eV. The basis of dimer fragments is already quite satisfactory with eigenvalue errors in the 10~meV range in the lowest part of the spectrum, and in the 30~meV range near the top of the valence band. The basis of trimer fragments gives excellent results with errors less than 1~meV in the lowest part of the spectrum and errors less than 10~meV near the top of the valence band.


Based on the results presented so far, we can conclude that the eigenfunctions of trimer fragments are an excellent basis for the representation of the Hamiltonian of the system. 
Since all the occupied states are calculated accurately, one can also imagine of using this approach for a full self-consistent DFT calculation without the use of the CPM.
However, for the present purpose, there is a strong interest to reduce the basis set as much as possible, since the reduction of the number of wave functions per fragment by a factor of $K$, reduces the time for their calculation by a factor of $K$, reduces the number of wave function overlaps and overlap integrals that need to be calculated by a factor of $K^2$ and reduces the computational time for the final diagonalization step by a factor of $K^3$.


\subsection{Choice of the number of basis states per fragment}

\begin{figure}[htbp]
 \centering
  \epsfig{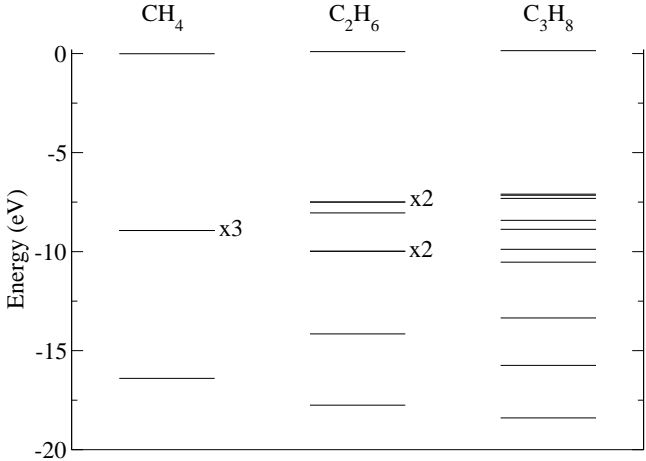}
  \caption{The eigenenergies of the monomer (methane), dimer (ethane) and trimer (propane). All occupied states and one unoccupied state are shown.}
 \label{fig:figA6}
 \end{figure}

\begin{figure}[htbp]
 \centering
  \epsfig{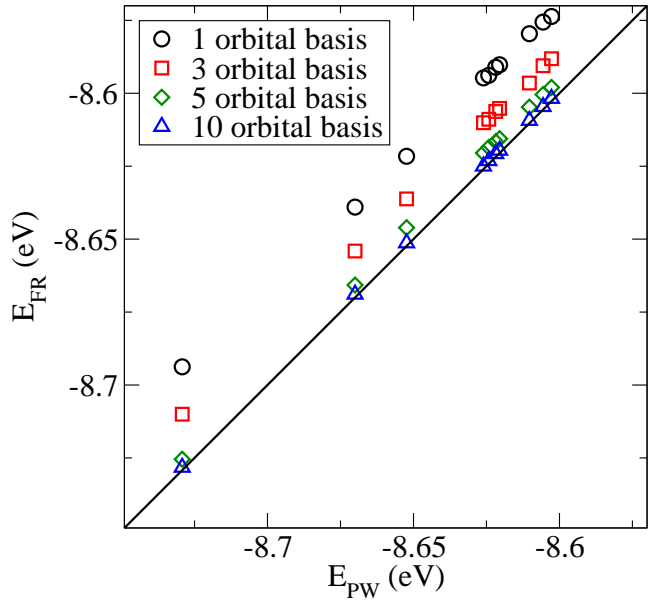}
  \caption{(Color online) The comparison of eigenenergies at the bottom of the valence band of amorphous alkane system calculated in the basis of trimer fragment wave functions $E_\mathrm{FR}$ and plane waves $E_\mathrm{PW}$. The number of the basis wave functions taken from each trimer is specified in the legend.}
 \label{fig:figA7}
 \end{figure}

The selection of fragment eigenstates which will be included in the basis set is based on physical intuition. For the lowest part of the spectrum, one expects that taking just the few lowest states of the trimer (whose eigenstates are shown in Fig.~\ref{fig:figA6}) should give quite accurate results. One can see from Fig.~\ref{fig:figA7} that even a single wave function per trimer gives quite satisfactory results, with a systematic error of the order of 30~meV only. This is not so surprising since the lowest state of the trimer is separated from the next one by about 3~eV (see Fig.~\ref{fig:figA6}) and therefore it is the only state that strongly contributes to the wave functions in the lowest part of the spectrum. With the inclusion of more states, the results converge towards the results obtained in plane wave basis. This is fully expected from the variational principle that states that the energy of any state formed from the finite basis set must be larger than the exact one and converges towards the exact one as the space spanned by the basis set is further increased.

\begin{figure}[htbp]
 \centering
  \epsfig{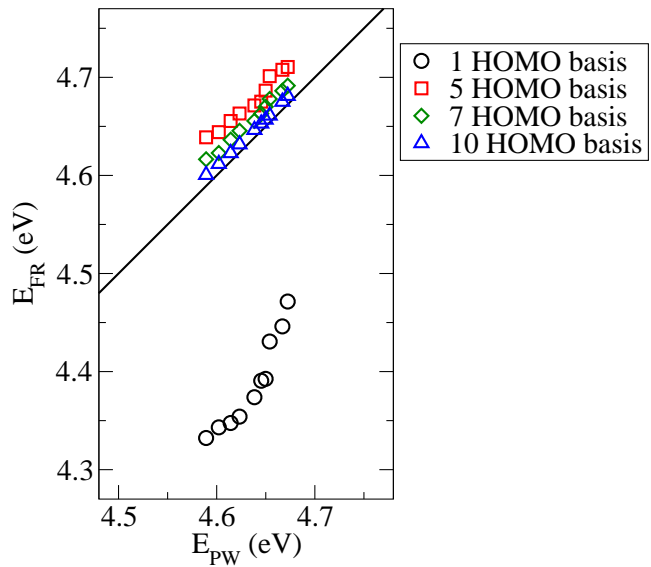}
  \caption{(Color online) The comparison of eigenenergies of amorphous alkane system calculated in the basis of trimer fragment wave functions $E_\mathrm{FR}$ and plane waves $E_\mathrm{PW}$. The number of the highest occupied basis wave functions taken from each trimer is specified in the legend. }
 \label{fig:figA8}
 \end{figure}

When one is interested in the part of the spectrum near the top of the valence band, one expects, based on physical intuition, that these states are formed from the highest occupied states of the fragment. However, in this case there is no exact principle that requires the states to converge (either from the top or bottom) toward the exact values as more highest occupied states of the fragment are added to the basis set. Indeed, we see from Fig.~\ref{fig:figA8} that the results obtained with one and five basis states per fragment are on the opposite side of the line with exact results. Furthermore, the results obtained with two, three or four highest occupied states per fragment, appear to have large basis set superposition errors and yield a completely different spectrum, where it is not even possible to correlate the eigenstates with the exact ones (these results are therefore not shown). The origin of such behavior comes from the energy level structure of the trimer fragment (Fig.~\ref{fig:figA6}). There are quite a few states near the highest occupied state and until all of them are included in the basis set, it is not possible to get a good description of the energy level structure.

{From these results, we may speculate about the general rule for the choice of basis states when one is interested in the states at the top of the valence band. One should make a cutoff based on energies of fragment states at the place where there is a substantial gap in their energies. However, it is difficult to predict in advance how many HOMOs are necessary. While in the case of alkanes at least 5 HOMOs are required to get reasonably accurate results (Fig.~\ref{fig:figA8}), in the case of thiophenes a single HOMO yields quite good results, as shown in Sec. \ref{sec:sec4}. Of course, the inclusion of all occupied states certainly leads to a good basis set. We find that it is often useful and practical to test the basis set convergence on some small systems (e.g., a single chain) where direct DFT calculation for
the whole system is possible, before using the current method to calculate large systems.} 

{It is also of substantial interest to determine which of the calculated states will be occupied and which not and consequently identify the Fermi level of the system. In the case when all occupied states of the fragments are used as basis set, it is trivial to occupy the states of the whole system based on the number of electrons in the system. In the case when only a few HOMOs of the fragments are taken into account, there is no such obvious procedure. Nevertheless, in practice we find it easy to recognize the HOMO - LUMO gap using the following procedure. We calculate $e_i=\langle i|H|i\rangle$, where $H$ is the Hamiltonian of the whole system and $|i\rangle$ the HOMO of the fragment. We then find the first gap in the density of states above $e_i$. That gap corresponds to the HOMO - LUMO gap. All states below that gap are then occupied, while the states above are empty.}

\subsection{Choice of the distance cutoff}
\begin{figure}[htbp]
 \centering
  \epsfig{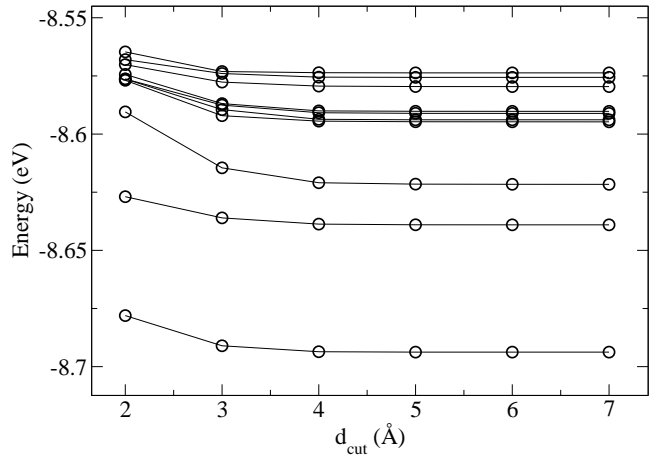}
  \caption{The dependence of the eigenenergies of amorphous alkane system in the lowest part of the valence spectrum on the cutoff distance $d_\mathrm{cut}$ between the fragments. The basis set with one wave function per fragment is used.}
 \label{fig:figA9}
 \end{figure}

An important factor that determines the accuracy of the calculation on the one hand and its speed on the other hand is the choice of pairs of fragments that are taken into account. We define the distance between fragments $j$ and $n$ as the minimal distance between an atom in fragment $j$ and an atom in fragment $n$. A pair of fragments $\{j,n\}$ is included in the calculation if the distance between them is smaller than some cutoff $d_\mathrm{cut}$. All the results presented so far have been obtained with $d_\mathrm{cut}$ (overcautiously) set to 7\AA. It is of interest to find the optimal value of $d_\mathrm{cut}$ which gives accurate eigenstates while minimizing the number of fragment pairs in the calculation. The dependence of the energies in the lowest part of the spectrum on $d_\mathrm{cut}$ is presented in Fig.~\ref{fig:figA9}, which shows that $d_\mathrm{cut}$=5\AA\ gives fully converged eigenstates.

\subsection{Dependence of computational time on system size}

\begin{figure}[htbp]
 \centering
  \epsfig{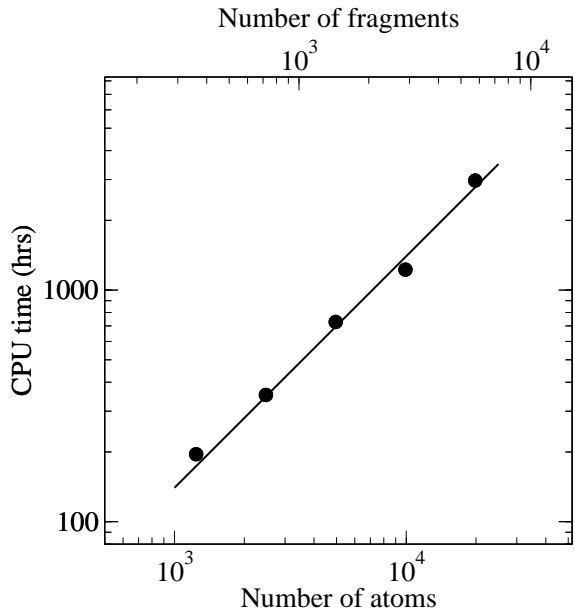}
  \caption{The dependence of the CPU time (defined as the wall clock time times the number of CPUs) on the size of the amorphous alkane system. The line is a fit to the $O(N)$ dependence. The calculations have been performed using one basis wave function per fragment. The number of CPUs in these calculations is typically of the order of 5000.}
 \label{fig:figA10}
 \end{figure}

\begin{figure}[htbp]
 \centering
  \epsfig{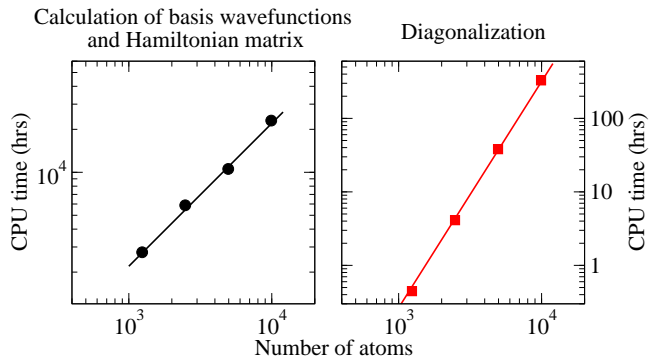}
  \caption{{(Color online) The dependence of the CPU time (defined as the wall clock time times the number of CPUs) on the size of the amorphous alkane system. The basis of ten wavefunctions per fragment is used in these calculations. The left panel shows the dependence of the time required for the calculation of the basis wavefunctions and Hamiltonian matrix elements [the line is a fit to the $O(N)$ dependence]. The right panel shows the time required for the solution of the generalized eigenvalue problem [the line is a fit to the $O(N^3)$ dependence]. The number of CPUs used for the calculation in the left panel was typically of the order of 5000, while in the right panel it ranged from 100 for the smallest system to 1600 for the largest system. This yields wallclock time of the order of several hours for the construction of Hamiltonian matrix and less than 10 minutes for its diagonalization.}}
 \label{fig:figXX}
 \end{figure}

The computational time for the {dominant part of the calculation in the} described methodology scales linearly with the size of the system in the size range considered in this paper, if the states in the fixed energy window are required, for the following reasons. The number of fragments is proportional to the size of the system, while the number of basis wave functions per fragment remains the same. Therefore, the time necessary to calculate all the basis wave functions is proportional to the number of fragments and consequently scales linearly with the size of the system. Furthermore, the number of fragment pairs with distance less than a certain predefined $d_\mathrm{cut}$ is also proportional to the number of atoms. For example, in the case of alkanes for $d_\mathrm{cut}=5$\AA, the number of pairs is approximately ten times larger than the number of atoms. {As a consequence, the CPU time for these two parts of the calculation scales linearly with the size of the system, as shown in Figs.~\ref{fig:figA10} and \ref{fig:figXX}.}
{The final diagonalization step, on the other hand, formally scales as $N^3$ with the size of the system. However, in the system size range that we consider this step takes a much smaller amount of time than the first two steps, as demonstrated in Fig. \ref{fig:figXX} (right panel). As a result, in this range of system dimensions, the total computational effort scales linearly with system size, as can be seen in Figs.~\ref{fig:figA10} and \ref{fig:figXX}.}

\section{Application to a conjugated polymer system}\label{sec:sec4}

The main motivation behind the development of this methodology was the lack of appropriate methods for the efficient calculation of the electronic structure of disordered conjugated polymers, where large supercells are required to provide insight into the physical properties of the system. Therefore, in this section, we test the applicability of the method to the calculation of hole states in P3HT, a widely studied polymer for applications in organic electronics.

We compare the eigenenergies obtained using OFM with the ones obtained by diagonalizing the CPM Hamiltonian using the plane wave basis set with kinetic energy cutoff of 60~Ry (which is done using the ESCAN code). For this test, we consider the system of 5 P3HT chains, each one being 20 thiophene rings long (which makes 2510 atoms altogether). The atomic structure of the system was generated from classical MD, using a simulated annealing procedure, as in our previous work.\cite{jpcb113-409} We make a comparison for ten different random realizations of the system, differing by initial conditions in MD simulation. We choose the basis of overlapping trimer fragments. In the fragments the side hexyl chains have been replaced by propyl chains. This replacement is motivated by the well known fact that wave functions in the region near the band edge that determine the electronic properties are localized on the main chain and not the alkyl side chains. Such a division into fragments would certainly not be sufficient to describe the part of electronic spectrum where the electronic states stemming from alkyl chains contribute to the density of states. That region is however far from the band gap region and is not of any physical interest. We further note that such a replacement by no means implies that the presence of hexyl side chains is ignored, in terms of their effects on the atomic structure of the system and the electrostatic potential in the full system Hamiltonian $H$ as constructed using the CPM. 

\begin{figure}[htbp]
 \centering
  \epsfig{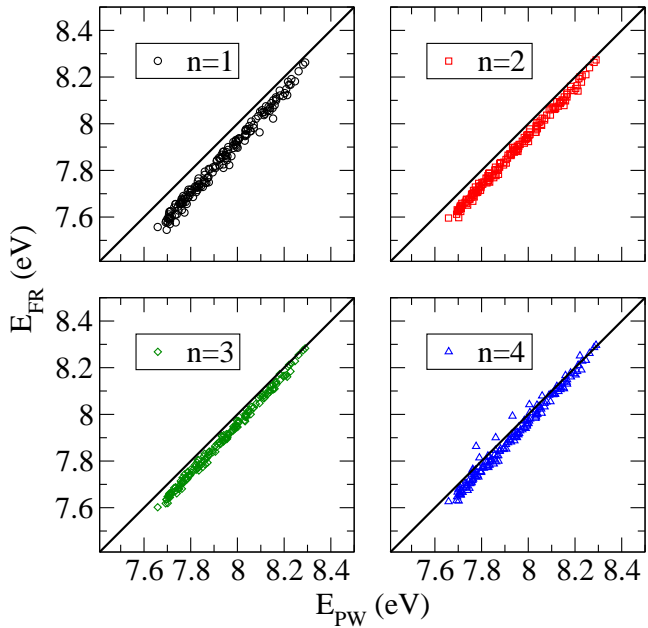}
  \caption{(Color online) The comparison of eigenenergies of the amorphous P3HT system obtained using the diagonalization of the Hamiltonian by the OFM $E_\mathrm{FR}$ and in the plane wave basis $E_\mathrm{PW}$. The straight line is given as a guide to the eye.}
 \label{fig:figB1}
 \end{figure}

We performed the test for different sizes of the basis set, consisting of $n$ top HOMOs of each fragment (where $n\in\{1,2,3,4\}$). The results gathered from all ten random realizations are presented in Fig.~\ref{fig:figB1}. The results obtained with $n=1$ are already quite accurate. The eigenenergy error for the states closest to the top of the valence band is of the order of 30~meV and increases to 120~meV as one goes 0.7~eV further away. The results are the most accurate for $n=3$ when the eigenenergy error is in the 10-50meV range. One should note that there is no exact principle that requires the eigenenergies to converge towards the "exact" ones as the basis set is increased and therefore there is no guarantee that a larger basis set would improve the results. Indeed, we find that for $n=4$ the results become worse than for $n=3$ (which can be evidenced by a larger dispersion of points and the presence of points both below and above the line). Finally, as one goes beyond $n=4$ certain states enter the band gap region and it becomes impossible even to establish a correspondence between the eigenstates from plane wave calculation with eigenstates from OFM calculation. 

{We would like to point out that our methodology strongly reduces the size of the basis set needed to represent the Hamiltonian of the system and for that reason makes the diagonalization part of the calculation the least demanding one. In the case of 2510 atom P3HT system, the basis of top three HOMOs per fragment consists of 270 elements and yields eigenenergies with errors in the 10-50 meV range. On the other hand, if the same system were considered using some typical basis of Gaussian orbitals, such as 6-31G$^*$, the size of the basis set would be 19720. In the case of alkanes, the gain is somewhat smaller. In our method with 5 HOMOs per fragment, that yields eigenenergy errors below 50~meV, we use 1800 basis wavefunctions to represent the 1240 atom alkane system. On the other hand, the 6-31G$^*$ basis set for the same system consists of 7680 wavefunctions.}

\section{Conclusion and outlook}
We have introduced the OFM for the calculation of eigenstates of the single-particle Hamiltonian. The method is based on the partitioning of the system into mutually overlapping fragments, the representation of the Hamiltonian in the basis of eigenstates of these fragments and the diagonalization of the obtained generalized eigenvalue problem. We have illustrated the method by applying it to find the eigenstates of the Hamiltonian of organic polymers obtained from the CPM. The method is expected to be more general -- it would be very interesting to test the method in other systems, such as for example inorganic nanostructures, inorganic alloys or any other organic structures - either ordered or disordered. The method is expected to be especially useful for understanding the properties of electronic states in the near-band gap tail of the density of states of statically or dynamically disordered systems, where large statistics is necessary to get reliable information. In this kind of systems the method would provide detailed information about the density of states in the tail, the wave function localization properties and consequently the electronic transport in the system. Furthermore, the method directly yields a parametrization of the Hamiltonian in a localized basis set and as such can be used as a starting point to build simple, but insightful tight-binding models of disordered systems. Finally, the method is naturally parallelizable and can make excellent use of parallel computing architectures, which have become the dominant paradigm in modern computing.

\section{Acknowledgments}

This work was supported by the DMS/BES/SC of the U.S. Department of Energy under Contract No. DE-AC02-05CH11231. It used the resources of National Energy Research Scientific Computing Center (NERSC) and the INCITE project allocations within the National Center for Computational Sciences (NCCS). In the last stages of this work, NV was supported by the Ministry of Science and Technological Development of the Republic of Serbia, under project No. ON171017, and the European Commission under EU FP7 projects PRACE-1IP, HP-SEE and EGI-InSPIRE.

\bibliographystyle{h-physrev3}

\end{document}